\documentclass{iaus}
\usepackage{graphicx}

\title[New halos around southern planetary nebulae] 
{Newly discovered halos and outer features around southern planetary nebulae
}

\author[D.\,J.~Frew, I.\,S.~Boji\v{c}i\'c \& Q.\,A.~Parker]   
{D.\,J.~Frew$^{1,2}$, I.\,S.~Boji\v{c}i\'c$^{1,2,3}$ and Q.\,A.~Parker$^{1,2,3}$}

\affiliation{$^1$Department of Physics and Astronomy, Macquarie University, Sydney, NSW 2109, Australia \\[\affilskip]
$^2$Macquarie University Research Centre in Astronomy, Astrophysics \& Astrophotonics\\[\affilskip]
$^3$Australian Astronomical Observatory, PO Box 296, Epping, NSW
1710, Australia\\email: {\tt david.frew@mq.edu.au}}

\pubyear{2012}
\volume{283}  
\pagerange{119--126}
 \jname{Planetary Nebulae: an Eye to the Future} \editors{Arturo Manchado, Letizia Stanghellini \& Detlef Sch\"onberner, eds.}
\begin{document}

\maketitle

\begin{abstract}
We have used the SuperCOSMOS H$\alpha$ Survey to look for faint outer structures such as halos, ansae and jets around known planetary nebulae across 4000 square degrees of the southern Milky Way.  Our search will contribute to  a more accurate census of these features in the Galactic PN population.  Candidate common-envelope PNe have also been identified on the basis of their microstructures.  We also intend to determine more reliable distances for these PNe, which should allow a much better statistical basis for the post-AGB total mass budget. Our survey offers fresh scope to address this important issue.
\keywords{Planetary nebulae: general --- ISM: general  --- stars: AGB and post-AGB --- stars: evolution}
\end{abstract}

\firstsection 
\section{Introduction}

One of the longest standing astrophysical problems is the exact relationship between the birth mass of stars and the mass left when they die. Unequal estimates of these masses have led to the planetary nebula ``missing mass''problem.  A typical PN progenitor star of mass 2.5\,$M_{\odot}$ sheds its envelope at the end of the AGB phase in a terminal superwind.  However, most PNe have an ionized mass of only 0.2--0.6\,$M_{\odot}$ while the residual white dwarf may be only 0.6\,$M_{\odot}$.  Until this mass discrepancy is reconciled, the gaps in our knowledge of the late evolution of intermediate-mass stars  impacts on our ability to model the evolution and chemical enrichment of our own Galaxy.

A typical AGB halo provides a visible fossil record of the AGB mass-loss history (e.g. Sandin et al. 2010), but has a mean H$\alpha$ brightness $\sim$10$^{3}\times$ fainter than the main PN shell, so a deep survey is needed.   Other complementary wavebands also provide evidence for the ubiquity of detectable haloes (see Chu, these proceedings), at least around young to middle-aged PNe.  For this optical study, we classify halos into the following groups: (i) AGB halos, (ii)  recombination halos, produced when the luminosity of the ionizing star drops rapidly (Corradi et al. 2003), and (iii) ISM halos surrounding optically thin PNe, a feature not fully appreciated in the literature.   Additional examples of the various halo types are required to allow a better statistical study of their distribution and properties.

\section{Methodology and Results}

\begin{figure}[tb]
\begin{center}
\includegraphics[width=5.3in ]
{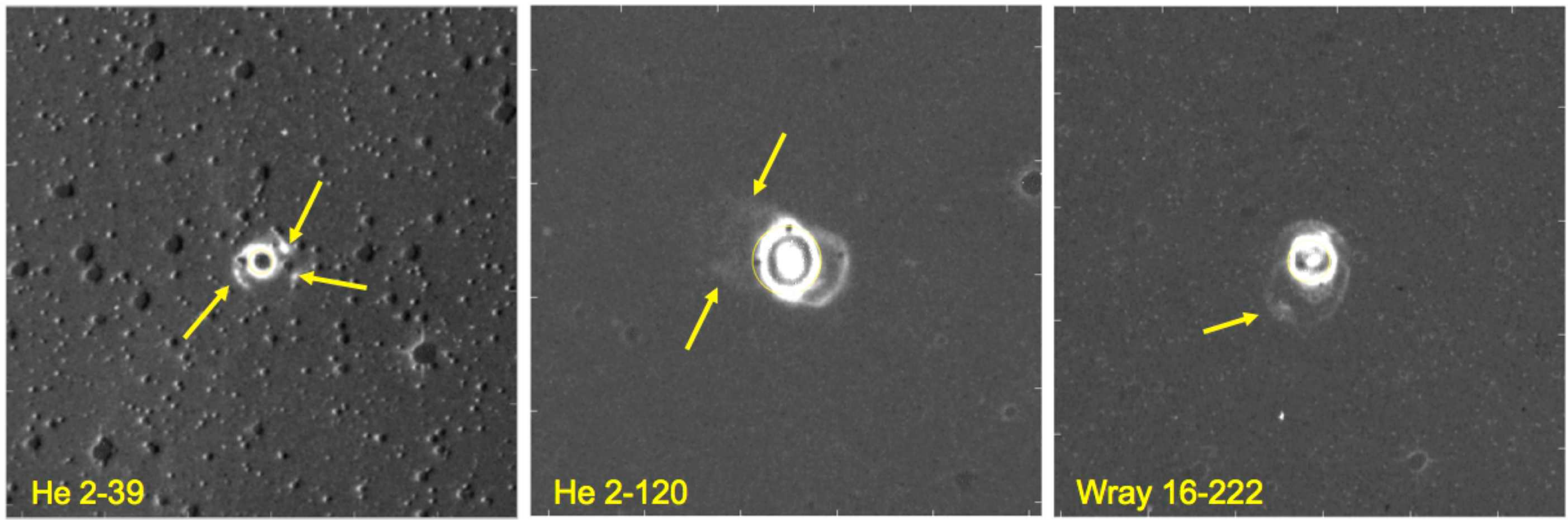}
\end{center}
\caption{SHS H$\alpha$/$R$ quotient images showing newly identified structures around three PNe.} 
\label{fig_halo}
\end{figure}

In order to improve the statistics of AGB halos and other external structures in the Galactic PN population, we have undertaken an extensive, systematic search around all known PNe using the SuperCOSMOS H$\alpha$ Survey (SHS), a deep, high-resolution H$\alpha$+[N\textsc{ii}] survey of the southern Galactic plane (Parker et al. 2005).   This study builds on unpublished work undertaken over the last decade from the original photographic films of the AAO/UKST H$\alpha$ Survey (e.g. Parker 2000).  
We examined PNe from the following catalogues:  Acker et al. (1992, 1996), Kohoutek (2001), the MASH catalogues (Parker et al. 2006; Miszalski et al. 2008), plus a significant number  from our own unpublished database; the only criterion was that the PN was located within the SHS coverage.  Quotient fits images (H$\alpha$/ broadband $R$) were generated for each PN in the survey area, of sufficient size to ensure that faint outer structures could be readily found.  The images were generated using the IRAF task $imarith$ before converting them to png format using the APLpy library (the Astronomical Plotting Library in Python).  Each image was linearly stretched using the default minimum and maximum pixel values (0.25\% and 99.75\%).  This allowed for rapid examination of the images in common image viewing software.  Some examples of newly identified outer features are shown in Figure~\ref{fig_halo}.

We have also noted any external low ionization structures (LIS; Gon\c{c}alves et al. 2001) such as fliers, ansae and jets around PNe in the survey zone. Such features are often seen in post-common envelope (CE) PNe (De Marco 2009; Miszalski et al. 2009), though it is unclear if LIS are also present in PNe that have not passed through a CE phase.   Additionally, faint extended lobes and irregular halos were found around a high fraction of Type I bipolar PNe (Frew et al. 2006),  which we attribute to the higher initial mass of their progenitor stars.  Lastly, we checked the nature of doubtful objects (Frew \& Parker 2010) as part of the search (see also Boissay et al., these proceedings).
Once this study is complete, a full catalogue of halos and outer structures will be published.

\begin{thebibliography}{}
\bibitem[Acker \etal\ (1992)]{Acker_etal92}{Acker, A., et al.} 1992, \textit{Strasbourg-ESO Catalogue of Galactic Planetary Nebulae}, Strasbourg
\bibitem[Acker \etal\ (1996)]{Acker_etal96}{Acker, A., et al.} 1996, \textit{Supplement to the Strasbourg-ESO Catalogue}, Strasbourg
\bibitem[Corradi \etal\ (2003)]{Corradi_etal03}{Corradi, R.L.M., Sch\"onberner, D., Steffen, M. \& Perinotto, M.} 2003, \textit{MNRAS}, 340, 417
\bibitem[De Marco (2009)]{dem_09}{De Marco, O.} 2009, \textit{PASP}, 121, 316
\bibitem[Frew \& Parker (2010)]{Frew_Parker10}{Frew, D.J. \& Parker, Q.A.} 2010, \textit{PASA}, 27, 129
\bibitem[Frew \etal\ (2006)]{Frew_etal06}{Frew, D.J., Parker, Q.A. \& Russeil, D.} 2006, \textit{MNRAS}, 372, 1081
\bibitem[Gon\c{c}alves \etal\ (2003)]{Goncalves_etal03}{Gon\c{c}alves, D., Corradi, R.L.M. \& Mampaso, A.} 2003, \textit{ApJ}, 547, 302
\bibitem[Kohoutek (2001)]{Kohoutek_01}{Kohoutek, L.} 2001, \textit{A\&A}, 378, 843
\bibitem[Miszalski \etal\ (2008)]{Miszalski_etal08}{Miszalski, B., Parker, Q.A., Acker, A., et al.} 2008, \textit{MNRAS}, 384, 525
\bibitem[Miszalski \etal\ (2009)]{Miszalski_etal09}{Miszalski, B., Acker, A.,  Parker, Q.A. \& Moffat, A.F.J.} 2009,  \textit{A\&A}, 505, 249
\bibitem[Parker (2000)]{Parker_00}{Parker, Q.A.} 2000, \textit{AAO Newsletter}, 94, 9
\bibitem[Parker \etal\ (2005)]{Parker_etal05}{Parker, Q.A., Phillipps, S., Pierce, M.J., et al.} 2005,  \textit{MNRAS}, 362, 689
\bibitem[Parker \etal\ (2006)]{Parker_etal06}{Parker, Q.A., Acker, A., Frew, D.J., et al.} 2006,  \textit{MNRAS}, 373, 79
\bibitem[Sandin \etal\ (2010)]{Sandin_etal10}{Sandin, C., Roth, M. M. \& Sch\"onberner, D.} 2010, \textit{PASA}, 27, 214


\end{thebibliography}
\end{document}